\documentclass[journal,12pt,draftclsnofoot,onecolumn]{IEEEtran}
\usepackage{mathrsfs}
\usepackage{cite}
\usepackage{amsmath}
\usepackage{amssymb}
\usepackage{stfloats}
\usepackage{graphicx}
\usepackage{setspace}
\usepackage{xcolor}

\ifCLASSOPTIONcompsoc
\usepackage[tight,normalsize,sf,SF]{subfigure}
\else
\usepackage[tight,footnotesize]{subfigure}
\fi

\begin{document}
\title{Design and Analysis of Downlink Channel Estimation Based on Parametric Model for Massive MIMO in FDD Systems}

\author{Yinsheng~Liu, Yinjun Liu, Qimei Cui, Riku J$\ddot{\text{a}}$ntti.

\thanks{Yinsheng Liu is with School of Computer Science and Information Technology and State Key Laboratory of Rail Traffic Control and Safety, Beijing Jiaotong University, Beijing 100044, China, e-mail: ys.liu@bjtu.edu.cn.}
\thanks{Yinjun Liu and Qimei Cui are with National Engineering Laboratory for Mobile Network Security,
Beijing University of Posts and Telecommunications, Beijing, 100876, China, email:liuyinjunbupt@gmail.com, cuiqimei@bupt.edu.cn.}
\thanks{Riku J$\ddot{\text{a}}$ntti is with the department of Communications and Networking at Aalto University School of Electrical Engineering, Finland, email: riku.jantti@aalto.fi.}
}

\maketitle
\doublespacing

\begin{abstract}
This paper investigates downlink channel estimation in frequency-division duplex (FDD)-based massive multiple-input multiple-output (MIMO) systems. To reduce the overhead of downlink channel estimation and uplink feedback in FDD systems, cascaded precoding has been used in massive MIMO such that only a low-dimensional effective channel needs to be estimated and fed back. On the other hand, traditional channel estimations can hardly achieve the minimum mean-square-error (MMSE) performance due to lack of the a priori knowledge of the channels. In this paper, we design and analyze a strategy for downlink channel estimation based on the parametric model in massive MIMO with cascaded precoding. For a parametric model, channel frequency responses are expressed using the path delays and the associated complex amplitudes. The path delays of uplink channels are first estimated and quantized at the base station, then fed forward to the user equipment (UE) through a dedicated feedforward link. In this manner, the UE can obtain the a priori knowledge of the downlink channel in advance since it has been demonstrated that the downlink and the uplink channels can have identical path delays. Our analysis and simulation results show that the proposed approach can achieve near-MMSE performance.
\end{abstract}


\newpage
\section{Introduction}
As a promising technique for the next generation cellular systems, massive multiple-input multiple-output (MIMO) has gained a lot attention recently \cite{FRusek}. By installing a huge number of antennas at the base station (BS), massive MIMO can significantly increase the spectrum- and energy-efficiencies of wireless networks \cite{HQNgo,EGLarsson}. \par

In massive MIMO systems, accurate downlink channel state information (CSI) is required at the user equipment (UE) for demodulation and the BS for precoding \cite{EGLarsson}. However, due to the large overhead caused by the huge number of antennas, both the estimation and the feedback of downlink CSI in frequency-division duplex (FDD)-based massive MIMO systems are not as easy as in regular MIMO systems. To address this issue, the spatial correlation of the channels corresponding to different antennas has been exploited to develop a cascaded precoding \cite{AAdhikary}, where we only need to deal with a low-dimensional effective channel such that traditional channel estimation and limited feedback can still be used \cite{YLiu_survey,DJLove}. The spatial correlation of the channels has also been used in \cite{JChoi2,AJDuly}, where a closed-loop training technique is used to improve the performance iteratively. In addition to the spatial domain, the delay-domain sparsity of the sampled channel impulse response (CIR) is also exploited in \cite{ZGao2,ZGao1} such that the number of unknowns for downlink channel estimation can be reduced significantly.\par

On the other hand, traditional channel estimation can hardly achieve the minimum mean-square-error (MMSE) performance due to lack of the a priori knowledge of the channels. To improve the performance, a parametric model has been used in \cite{BYang} where the channel frequency respones (CFR) are expressed using the path delays and the associated complex amplitudes. By estimating the path delays and the complex amplitudes separately, the estimation accuracy can be greatly improved. In \cite{BYang}, estimation of signal parameters by rotational invariance technique (ESPRIT) has been used to estimate the path delays \cite{RRoy}, which requires a long symbol sequence to obtain the frequency-domain covariance matrix. Although some approaches have been proposed to reduce the sequence length \cite{MRRaghavendra2,YLiu_survey}, the need of the long symbol sequence still limits the application of parametric model based channel estimation in burst-type transmissions, such as the cellular systems. \par

In this paper, we will exploit the large number of antennas in massive MIMO systems and the reciprocity of the path delays between the downlink and the uplink to design a strategy for downlink channel estimation based on the parametric model in a massive MIMO system with cascaded precoding. On one hand, the long symbol sequence needed in traditional parametric model based approaches is unnecessary in massive MIMO systems. The frequency-domain covariance matrix can be estimated using the large number of antennas at the BS, and thus the path delays for the uplink channel can be obtained efficiently at the BS. On the other hand, in FDD systems, the separation between the downlink and the uplink frequencies are about $5\%$ of the center frequency \cite{3GPP}. Given such small frequency separation, the downlink and the uplink will have many common features such as the path delays \cite{AJPaulraj ,AJPaulraj2}. In this case, the uplink path delays estimated at the BS can be directly used as the downlink ones.\par

The above observations inspire us to design a strategy for downlink channel estimation in this paper. In the proposed strategy, the path delays of uplink channels are first estimated at the BS using the large number of antennas, then quantized and fed forward to the UE through a dedicated feedforward link. In this manner, the UE can obtain the knowledge of the downlink channel in advance since the downlink and uplink have identical path delays. Once the UE has the knowledge of the path delays, the effective CFR can be regenerated by estimating the associated complex amplitudes. Our analysis shows that the performance of channel estimation can be improved by increasing the accuracy of path delay estimation or using more quantization bits. Given sufficient quantization bits and accurate path delay estimation, the proposed strategy can achieve near-MMSE performance.\par

It should be highlighted that our approach in this paper can be also used in general downlink massive MIMO systems without cascaded precoding. However, since it is difficult to achieve efficient feedback in such systems, we focus on the cascaded precoding based massive MIMO system which is more realistic for practical realization.\par

The rest of this paper is organized as follows. The system model is presented in Section II. The channel estimation strategy is described in Section III, and analyzed in Section IV. Simulation results are shown in Section V, and conclusions are finally drawn in Section VI.

\section{System Model}

\begin{figure*}
  \centering
  \includegraphics[width=5in]{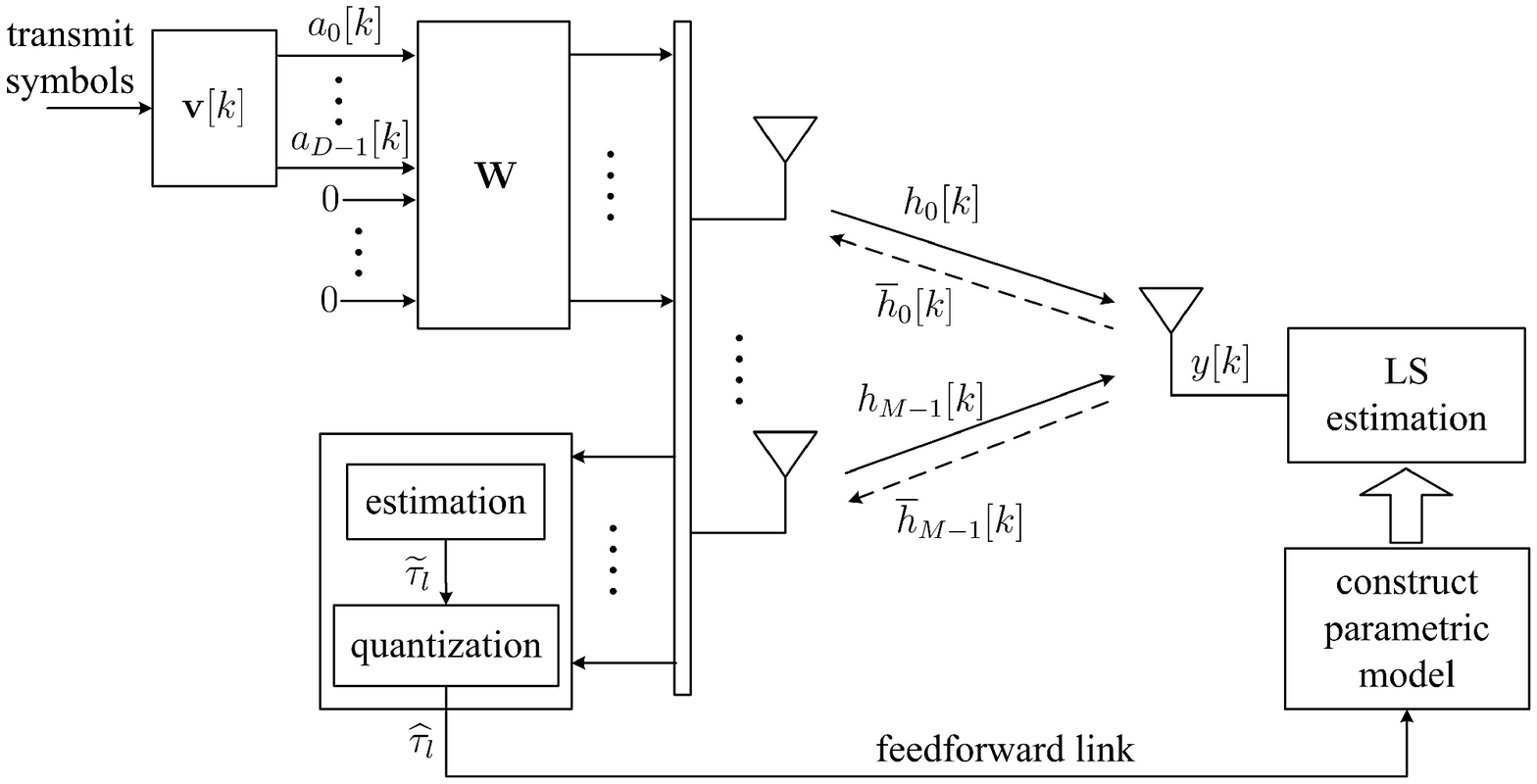}\\
  (a)\\
  \includegraphics[width=2.5in]{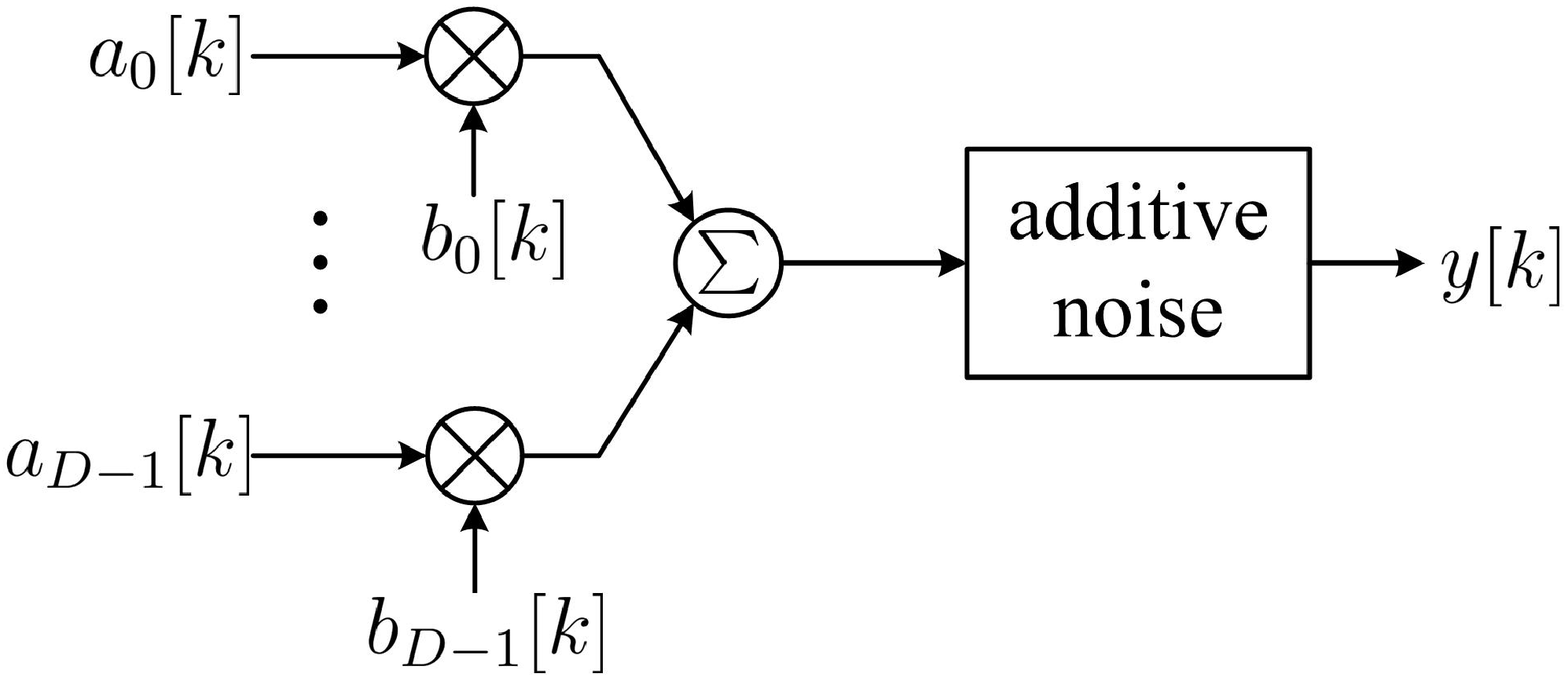}\\
  (b)
  \caption{A massive MIMO system with cascaded precoding in (a) where the path delays are estimated at the BS and then fed forward to the UE through a dedicated link, and (b) an effective downlink channel model.}\label{system}
\end{figure*}

As in Fig.~\ref{system} (a), we consider an orthogonal frequency division multiplexing (OFDM) with $K$ subcarriers in a massive MIMO system with $M$ antennas at the BS and a single antenna at the UE of interest. Only the single user case is considered in this paper although our approach can be also extended to the multiuser case. In Fig.~\ref{system} (a), $\mathbf{v}[k]\in\mathcal{C}^{D\times 1}$ denotes the outer precoder at the $k$-th subcarrier where $D$ is the size of the effective channel with $D\ll M$, and $\mathbf{W}\in\mathcal{C}^{M\times D}$ denotes the inner precoder \cite{AAdhikary}.\par

Denote $h_m[k]$ to be the dowlink CFR at the $k$-th subcarrier corresponding to the $m$-th antenna. In the presence of multipath propagation, $h_m[k]$ can be expressed as
\begin{align}\label{dchan2}
h_m[k]=\sum_{l=0}^{L-1}\alpha_m[l]e^{-j\frac{2\pi k}{T}\tau_l},
\end{align}
where $T$ is the OFDM symbol duration, $\tau_l$ denotes the delay of the $l$-th path, and $\alpha_m[l]$ denotes the complex amplitude of the $l$-th path at the $m$-th antenna. We assume the complex amplitudes are Gaussian distributed with zero mean and $\mathrm{E}(|\alpha_m[l]|^2)=\sigma_l^2$ with $\sum_{l=0}^{L-1}\sigma_l^2=1$. Furthermore, the complex amplitudes corresponding to different paths are assumed independently distributted in this paper.\par

When a large number of antennas are placed in a small area, the channels at different antennas will be correlated. Accordingly, the spatial correlation function can be defined as
\begin{align}\label{rs}
r_s[m]\triangleq\mathrm{E}(h_{m+n}[k]h_n^*[k]),
\end{align}
or in a matrix form as $\mathbf{R}_s=\{r_s[m-n]\}_{m,n=0}^{M-1}$. From \cite{AAdhikary}, the optimal inner precoder is given by $\mathbf{W}=\mathbf{U}_s^*$ where $\mathbf{U}_s=(\mathbf{u}_s[0],\cdots,\mathbf{u}_{s}[D-1])$ is composed of $D$ eigenvectors associated with the largest eigenvalues of downlink spatial covariance matrix, $\mathbf{R}_s$. In practical systems, the inner precoder can be obtained by exploiting the reciprocity of the spatial covarince matrices between the downlink and the uplink \cite{CSun,GBarriac}. In this paper, we assume the inner precoder is ideally known such that we can focus on the channel estimation with respect to the low-dimensional effective channel.\par

Denote $\mathbf{h}[k]=(h_0[k],\cdots,h_{M-1}[k])^{\mathrm{T}}$ to be a channel vector composed of the actual downlink CFRs from all the antennas at the $k$-th subcarrier. Then, given the optimal inner precoder, the effective downlink CFR at the $k$-th subcarrier over the $d$-th eigenvector can be expressed as
\begin{align}\label{1_0}
b_d[k]=\mathbf{u}^{\mathrm{H}}_s[d]\mathbf{h}[k].
\end{align}
In this case, the received signal at the $k$-th subcarrier over the effective channel can be expressed, from Fig.~\ref{system} (b), as
\begin{align}\label{signal2}
y[k]=\sum_{d=0}^{D-1}a_d[k]b_d[k]+z[k],
\end{align}
where $z[k]$ denotes the additive white Gaussian noise with zero mean and $\mathrm{E}(|z[k]|^2)=N_0$, and $a_d[k]$ denotes the frequency-domain training symbol over the $d$-th eigenvector. In this paper, we use training symbols with constant amplitudes and random phases, that is, $a_d[k]=e^{j\phi_d[k]}$. The phases are independently generated for different $d$'s and $k$'s with a uniform distribution in $[-\pi,\pi)$.\par

Theoretically, the downlink and the uplink channels should have the same path delays since the signals will travel the same distance in both ways \cite{YIWu,AYOlenko}. Measurement results in \cite{SSalous,SSalous1,SSalous2} have shown that the power delay profiles are indeed very similar for the downlink and the uplink. It is therefore reasonable to assume identical path delays for both downlink and uplink as in \cite{AJPaulraj ,AJPaulraj2,GGRaleigh,CSun}. Experimental results in \cite{NPromsuvana} have demonstrated that the assumption of identical path delays coincide with the practical measurements given small frequency separation between the downlink and the uplink.\par
When the downlink and the uplink channels have the same path delays, the uplink channel at the $k$-th subcarrier on the $m$-th antenna can be given as
\begin{align}\label{uchan}
\overline{h}_m[k]=\sum_{l=0}^{L-1}\overline{\alpha}_m[l]e^{-j\frac{2\pi k}{T}\tau_l},
\end{align}
where $\overline{\alpha}_m[l]$, with zero mean and $\mathrm{E}(|\overline{\alpha}_m[l]|^2)=\overline{\sigma}_l^2$, denotes the complex amplitude of the $l$-th path at the $m$-th antenna for the uplink. Similarly, the complex amplitudes corresponding to different paths for the uplink are also assumed independent.\par

Based on (\ref{dchan2}) and (\ref{uchan}), the path delays, $\tau_l$'s, for the downlink channel can be obtained at the BS through uplink channels since they have common path delays.

\section{Channel Estimation Strategy Based on Parametric Model}
In this section, we first develop the parametric model in massive MIMO systems with cascaded precoding. Then, we discuss estimation, quantization, and feedforward of the path delays. Finally, we present the least-square (LS) estimation of the associated complex amplitudes.

\subsection{Parametric Model in Massive MIMO with Cascaded Precoding}

For the parametric model, the CFR is represented using the path delays and the associated complex amplitude of each path \cite{BYang}. To obtain the parametric model in cascaded precoding based massive MIMO, denote $\boldsymbol{\alpha}[l]=(\alpha_0[l],\cdots,\alpha_{M-1}[l])^{\mathrm{T}}$ to be the vector composed of the actual complex amplitudes from all the antennas at the $l$-th path. Then, from (\ref{dchan2}) and (\ref{1_0}), the effective complex amplitude corresponding to the $l$-th path over the $d$-th eigenvector can be obtained as
\begin{align}\label{beta_exp}
\beta_d[l]=\mathbf{u}^{\mathrm{H}}_s[d]\boldsymbol{\alpha}[l].
\end{align}
Similar to (\ref{dchan2}), the effective CFR, $b_d[k]$, can be expressed as the Fourier transform of the effective CIR, that is
\begin{align}\label{CFR_CIR}
b_d[k]=\sum_{l=0}^{L-1}\beta_d[l]e^{-j\frac{2\pi k}{T}\tau_l}.
\end{align}
Therefore, the effective CFRs, $b_d[k]$'s, can be obtained by estimating the path delays, $\tau_l$'s, and the effective complex amplitudes, $\beta_d[l]$'s, respectively.\par

Substituting (\ref{CFR_CIR}) into (\ref{signal2}), the received signal based on the parametric model is rewritten as
\begin{align}\label{param_model}
y[k]=\sum_{d=0}^{D-1}a_d[k]\left(\sum_{l=0}^{L-1}\beta_d[l]e^{-j\frac{2\pi k}{T}\tau_l}\right)+z[k].
\end{align}
If taking all the subcarriers into account, (\ref{param_model}) can be rewritten in a matrix form as
\begin{align}\label{IO}
\mathbf{y}=\sum_{d=0}^{D-1}\mathbf{A}_d\mathbf{S}\boldsymbol{\beta}_d+\mathbf{z},
\end{align}
where $\mathbf{y}=(y[0],\cdots,y[K-1])^{\mathrm{T}}$, $\mathbf{A}_d=\mathrm{diag}\{a_d[k]\}_{k=0}^{K-1}$, $\boldsymbol{\beta}_d=(\beta_d[0],\cdots,\beta_d[L-1])^{\mathrm{T}}$, $\mathbf{z}=(z[0],\cdots,z[K-1])^{\mathrm{T}}$, and $\mathbf{S}=[\mathbf{s}(\tau_0),\cdots,\mathbf{s}(\tau_{L-1})]$ with $\mathbf{s}(\tau_l)=[1,e^{-j\frac{2\pi}{T}\tau_l},\cdots,e^{-j\frac{2\pi(K-1)}{T}\tau_l}]^{\mathrm{T}}$ indicating the frequency-domain steering vector. For a more tight form, (\ref{IO}) can be rewritten as
\begin{align}\label{LS_target}
\mathbf{y}=\mathbf{X}\boldsymbol{\beta}+\mathbf{z},
\end{align}
where $\mathbf{X}=[\mathbf{A}_0\mathbf{S},\cdots,\mathbf{A}_{D-1}\mathbf{S}]$ and
\begin{align}
\boldsymbol{\beta}=\left(
\begin{array}{c}
  \boldsymbol{\beta}_0 \\
  \vdots \\
  \boldsymbol{\beta}_{D-1}
\end{array}
\right).
\end{align}

\subsection{Path Delay: Estimation, Quantization, and Feedforward}
\subsubsection{Estimation}
Since the path delays are identical for the downlink and the uplink, the estimated path delays from the uplink can be directly used as the downlink ones. The subspace-based approach, which consists of two steps, can be used for the estimation of the uplink delays at the BS \cite{BYang}.\par
The first step is to estimate the uplink frequency-domain covariance matrix, which can be given as $\overline{\mathbf{R}}_f=\{\overline{r}_f[k-p]\}_{k,p=0}^{K-1}$ where $\overline{r}_f[k]\triangleq \mathrm{E}(\overline{h}_{m}[q+k]\overline{h}_m^*[q])$ denotes the corresponding correlation function. In massive MIMO systems, the frequency-domain covariance matrix can be estimated by averaging the uplink CFRs corresponding to different antennas at the BS,
\begin{align}
\widetilde{\overline{\mathbf{R}}}_f=\frac{1}{M}\sum_{m=0}^{M-1}\overline{\mathbf{h}}_m\overline{\mathbf{h}}_m^{\mathrm{H}},
\end{align}
where $\overline{\mathbf{h}}_m=(\overline{h}_m[0],\cdots,\overline{h}_m[K-1])^{\mathrm{T}}$ denotes the uplink channel vector, which can be estimated via uplink channel estimation \cite{YLiu_survey}. In this case, the long symbol sequence in \cite{BYang,MRRaghavendra2} is not required any more and the proposed strategy can be used in cellular systems.\par
For the second step, the ESPRIT algorithm can be used to obtain the estimation of the path delays as in \cite{BYang}. The procedure is the same with that in \cite{BYang} and thus not presented here.\par

The accuracy of path delay estimation can be measured by the variance, $\mathrm{E}(|\widetilde{\tau}_l-\tau_l|^2)$, where $\widetilde{\tau}_l$ denotes the corresponding estimated path delay. Although the estimation performance of ESPRIT algorithm can be improved by using more subcarriers and more antennas, it is in general difficult to obtain an analytical result \cite{FLi}, and we therefore use a simple notation
\begin{align}
\mathrm{E}(|\widetilde{\tau}_l-{\tau}_l|^2)={\sigma}^2,\label{delay_sta2}
\end{align}
as the performance metric of the path delay estimation, where ${\sigma}^2$ decreases as the increase of numbers of subcarriers or antennas \cite{FLi}.

\begin{figure}
  \centering
  \includegraphics[width=3in]{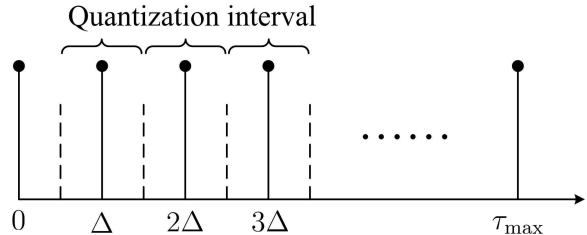}\\
  \caption{Quantization of the path delay.}\label{quantization}
\end{figure}

\subsubsection{Quantization}
The estimated path delay, $\widetilde{\tau}_l$, is then quantized to, $\widehat{\tau}_l$, so that it can be fed forward to the UE. In this paper, we use a simple uniform quantization as in Fig.~\ref{quantization}. If $B$ bits are used for the quantization of each path delay, then the quantization interval can be given by
\begin{align}
\Delta = \frac{\tau_{\mathrm{max}}}{2^B},
\end{align}
where $\tau_{\mathrm{max}}$ indicates the maximum delay. In practical systems, the duration of the cyclic prefix can be used instead if the maximum delay is unknown.\par
For uniform quantization, the quantization error, $\widehat{\tau}_l-\widetilde{\tau}_l$, can be viewed as a uniformly distributed noise with zero mean and \cite{BWidrow}
\begin{align}
\mathrm{E}(|\widehat{\tau}_l-\widetilde{\tau}_l|^2)=\frac{\Delta^2}{12}=\frac{\tau^2_{\mathrm{max}}}{12\cdot 4^{B}},\label{delay_sta1}
\end{align}
which shows that the quantization performance can be improved exponentially by using more quantization bits.


\subsubsection{Feedforward}
After the quantization, the path delays are fed forward to the UE through a dedicated feedforward link. Similar to traditional limited feedback \cite{DJLove}, we assume the feedforward link is error-free. On the other hand, the feedback delay in traditional limited feedback may deteriorate the system performance due to the time variation of wireless channels. In the scenario of this paper, however, the path delay depends on the surrounding scatters in typical wireless channels, which will not change for a relatively longer duration. It is therefore reasonable to assume that the path delays are constant and thus the feedback delay has no impact on the the proposed approach.

\subsection{LS Estimation of Complex Amplitudes}
Once the UE has the knowledge of the path delays, it only needs to estimate the effective complex amplitudes. From (\ref{LS_target}), the LS estimation of the effective complex amplitudes can be given by
\begin{align}\label{LS}
\widehat{\boldsymbol{\beta}}&=(\widehat{\mathbf{X}}^{\mathrm{H}}\widehat{\mathbf{X}})^{-1}\widehat{\mathbf{X}}^{\mathrm{H}}\mathbf{y},
\end{align}
where $\widehat{\mathbf{X}}$ is exactly the same with $\mathbf{X}$ except that the quantized delays, $\widehat{\tau}_l$'s, are used instead of the real ones. Note that if two or more path delays are too close to separate due to inaccurate path delay estimation or small number of quantization bits, we can view those inseparable path delays as a single one so that $\widehat{\mathbf{X}}^{\mathrm{H}}\widehat{\mathbf{X}}$ in (\ref{LS}) is always with full rank.\par

Given the quantized path delays and estimated effective complex ampltitudes, the estimated effective CFR can be regenerated, similar to (\ref{CFR_CIR}), as
\begin{align}
\widehat{b}_d[k]=\sum_{l=0}^{L-1}\widehat{\beta}_d[l]e^{-j\frac{2\pi k}{T}\widehat{\tau}_l}.
\end{align}


\section{Performance Analysis}
In this section, we will first analyze the performance of the proposed channel estimation strategy in Section III. Then, a comparison with the MMSE estimator will be shown.

\subsection{Mean-Square-Error (MSE) for Channel Estimation}

The MSE for channel estimation can be defined as
\begin{align}\label{chan_mse}
\mathrm{MSE}\triangleq \sum_{d=0}^{D-1}\sum_{k=0}^{K-1}\mathrm{E}(|\widehat{b}_d[k]-b_d[k]|_2^2),
\end{align}
where the expectation is with respect to the channels and the additive noise. When the number of subcarriers is large enough as in most systems, we have \cite{MViberg}
\begin{align}\label{orth}
\frac{1}{K}\mathbf{s}^{\mathrm{H}}(\tau_l)\mathbf{s}(\tau_p)&=\mathrm{sinc}\left[\frac{\pi (\tau_{l}-{\tau}_{p})K}{T}\right]e^{j\frac{\pi (K-1)}{T}(\tau_{l}-{\tau}_{p})}\nonumber\\
&\approx \left\{
           \begin{array}{ll}
             1, & \tau_l=\tau_p \\
             0, & \tau_l\neq\tau_p
           \end{array}.
         \right.
\end{align}
Using the relation above together with Appendix A, we have
\begin{align}\label{mse2}
\mathrm{MSE}=K\sum_{l=0}^{L-1}\mathrm{Tr}\{\mathbf{U}_s^{\mathrm{H}}\mathbf{R}_{s,l}\mathbf{U}_s\}\left\{1-\mathrm{sinc}^2\left[\frac{\pi (\widehat{\tau}_{l}-\tau_l)K}{T}\right]\right\}+N_0 LD,
\end{align}
where $\mathbf{R}_{s,l}=\{r_{s,l}[m-n]\}_{m,n=0}^{M-1}$ denotes the dowlink spatial covariance matrix caused by the subpaths inside the $l$-th path with $r_{s,l}[m]$ indicating the corresponding correlation function,
\begin{align}
r_{s,l}[m]\triangleq \mathrm{E}(\alpha_{n+m}[l]\alpha_n^*[l]),
\end{align}
and we can obtain, from (\ref{rs}), that $\mathbf{R}_s=\sum_{l=0}^{L-1}\mathbf{R}_{s,l}$.\par

In (\ref{mse2}), the overall error, $\widehat{\tau}_l-\tau_l$, is composed of a quantization error and an estimation error,
$\widehat{\tau}_l-\tau_l=(\widehat{\tau}_l-\widetilde{\tau}_l)+(\widetilde{\tau}_l-\tau_l)$. When the antenna number is large, the estimation error can be very small \cite{FLi}. Similarly, the quantization error can also be very small if we assume the number of quantization bits is sufficiently large. Under this situation, $\widehat{\tau}_l$ and ${\tau}_l$ will be very close and thus
\begin{align}\label{sinc}
\mathrm{sinc}\left[\frac{\pi(\widehat{\tau}_l-{\tau}_l)K}{T}\right]\approx 1-\frac{\pi^2K^2}{6T^2}(\widehat{\tau}_l-{\tau}_l)^2.
\end{align}
Using (\ref{sinc}), (\ref{mse2}) can be rewritten as
\begin{align}\label{mse3_0}
\overline{\mathrm{MSE}}&=\frac{\pi^2K^3}{3T^2}\sum_{l=0}^{L-1}\mathrm{Tr}\{\mathbf{U}_s^{\mathrm{H}}\mathbf{R}_{s,l}\mathbf{U}_s\}\mathrm{E}(|\widehat{\tau}_l-{\tau}_l|^2)
+N_0 LD\nonumber\\
&=\frac{\pi^2K^3}{3T^2}\sum_{l=0}^{L-1}\mathrm{Tr}\{\mathbf{U}_s^{\mathrm{H}}\mathbf{R}_{s,l}\mathbf{U}_s\}\left[\mathrm{E}(|\widehat{\tau}_l-\widetilde{\tau}_l|^2) + \mathrm{E}(|\widetilde{\tau}_l-{\tau}_l|^2)\right]
+N_0 LD,
\end{align}
where we have ignored the high-order error term. The second equation in (\ref{mse3_0}) is due to the assumption that the estimation error and the quantization error are independent. Substituting (\ref{delay_sta2}) and (\ref{delay_sta1}) into (\ref{mse3_0}), we obtain
\begin{align}\label{mse4}
\overline{\mathrm{MSE}}=\frac{\pi^2K^3}{3T^2}\mathrm{Tr}\{\mathbf{U}_s^{\mathrm{H}}\mathbf{R}_s\mathbf{U}_s\}\cdot\left(\frac{\tau^2_{\mathrm{max}}}{12\cdot 4^{B}}+{\sigma}^2\right)+N_0 LD.
\end{align}
Equation (\ref{mse4}) shows that the performance of channel estimation can be improved by using more quantization bits or increasing the accuracy of path delay estimation. In an extreme case where the path delay estimation is ideal and $B\rightarrow \infty$, the average MSE is further reduced to
\begin{align}\label{mse5_0}
\overline{\mathrm{MSE}}=N_0 LD,
\end{align}
which is proportional to the number of effective complex amplitudes.\par

For more insights, the term, $\mathrm{Tr}\{\mathbf{U}_s^{\mathrm{H}}\mathbf{R}_s\mathbf{U}_s\}$, in (\ref{mse4}) can be rewritten as
\begin{align}\label{insight0}
\mathrm{Tr}\{\mathbf{U}_s^{\mathrm{H}}\mathbf{R}_s\mathbf{U}_s\}=L \cdot g_D\left(\frac{1}{L}\mathbf{R}_s\right),
\end{align}
where $g_D(\cdot)$ denotes the sum of the $D$ largest eigenvalues of a given matrix, that is
\begin{align}
g_D\left(\frac{1}{L}\mathbf{R}_s\right)=\sum_{d=0}^{D-1}\lambda_d\left(\frac{1}{L}\mathbf{R}_s\right),
\end{align}
where $\lambda_d(\cdot)$ denotes the $d$-th largest eigenvalue of a given matrix. From \cite{ABTal,WWATKINS}, $g_D(\cdot)$ is a convex function, such that
\begin{align}\label{insight1}
g_D\left(\frac{1}{L}\mathbf{R}_s\right)=g_D\left(\frac{1}{L}\sum_{l=0}^{L-1}\mathbf{R}_{s,l}\right)\leq \frac{1}{L}\sum_{l=0}^{L-1}g_D(\mathbf{R}_{s,l}),
\end{align}
where the equation holds when $\mathbf{R}_{s,l}=\frac{1}{L}\mathbf{R}_s$ for all $l$'s. In this case, we have
\begin{align}
\mathrm{E}\{|\beta_d[l]|^2\}=\frac{1}{L}\mathbf{u}_s^{\mathrm{H}}[d]\mathbf{R}_s\mathbf{u}_s[d],
\end{align}
and thus $\mathrm{E}\{|\beta_d[l]|^2\}$'s are constant for all $l$'s. If assuming the $D$ largest eigenvalues can capture all the power of $\mathbf{R}_{s,l}$, then
\begin{align}
g_D(\mathbf{R}_{s,l})=\mathrm{Tr}\{\mathbf{R}_{s,l}\}=M\sigma_{l}^2,
\end{align}
and therefore (\ref{insight1}) can be rewritten as
\begin{align}\label{insight2}
g_D\left(\frac{1}{L}\mathbf{R}_s\right)\leq \frac{M}{L}\sum_{l=0}^{L-1}\sigma_{l}^2=\frac{M}{L}.
\end{align}
As a result, by substituting (\ref{insight2}) into (\ref{insight0}), we can obtain
\begin{align}
\mathrm{Tr}\{\mathbf{U}_s^{\mathrm{H}}\mathbf{R}_s\mathbf{U}_s\}\leq M,
\end{align}
which suggests that an effective channel with equal powers among the paths gives the worst performance
\begin{align}
\overline{\mathrm{MSE}}=\frac{\pi^2K^3 M}{3T^2}\left(\frac{\tau^2_{\mathrm{max}}}{12\cdot 4^{B}}+{\sigma}^2\right)+N_0 LD.
\end{align}

\subsection{Comparison with MMSE Estimator}
It is necessary to make a comparison with the MMSE estimator since it can achieve the best performance. If using the MMSE estimation for the effective CFR in (\ref{signal2}), the associated MSE can be given as \cite{SHaykin}
\begin{align}\label{mse_mmse}
{\mathrm{MSE}_{\mathrm{MMSE}}}=\mathrm{Tr}\left\{\left(\mathbf{R}_b^{-1}+\frac{1}{N_0}\mathbf{A}^{\mathrm{H}}\mathbf{A}\right)^{-1}\right\},
\end{align}
where $\mathbf{A}=(\mathbf{A}_0,\cdots,\mathbf{A}_{D-1})$ and
\begin{align}
\mathbf{R}_{b}=\left(
\begin{array}{ccc}
  \mathrm{E}(\mathbf{b}_{0}\mathbf{b}_{0}^{\mathrm{H}}) & \cdots & \mathrm{E}(\mathbf{b}_0\mathbf{b}_{D-1}^{\mathrm{H}}) \\
  \vdots & \ddots & \vdots \\
  \mathrm{E}(\mathbf{b}_{D-1}\mathbf{b}_{0}^{\mathrm{H}}) & \cdots &  \mathrm{E}(\mathbf{b}_{D-1}\mathbf{b}_{D-1}^{\mathrm{H}})
\end{array}
\right),
\end{align}
with $\mathbf{b}_d=(b_d[0],\cdots,b_d[K-1])^{\mathrm{T}}$.
Similar to \cite{OEdfors}, we use $\mathrm{E}(\mathbf{A}^{\mathrm{H}}\mathbf{A})=\mathbf{I}$ to replace $\mathbf{A}^{\mathrm{H}}\mathbf{A}$ in (\ref{mse_mmse}) such that the analysis can be greatly simplified. Then, from Appendix B, the MSE in (\ref{mse_mmse}) can be given as
\begin{align}\label{mse_mmse2}
\mathrm{MSE}_{\mathrm{MMSE}}=N_0\sum_{i=0}^{LD-1}\frac{\lambda_b[i]}{\lambda_b[i]+N_0},
\end{align}
where $\lambda_b[i]$ indicates the $i$-th eigenvalue of $\mathbf{R}_b$.\par

Since $\lambda_b[i](\lambda_b[i]+N_0)^{-1}<1$, we can obtain that
\begin{align}
\mathrm{MSE}_{\mathrm{MMSE}}<N_0L D,
\end{align}
which means the MMSE estimator is always better than the proposed strategy. However, when the signal-to-noise ratio (SNR) is large enough, $\lambda_b[i](\lambda_b[i]+N_0)^{-1}\approx 1$ and thus
\begin{align}
\mathrm{MSE}_{\mathrm{MMSE}}\approx N_0LD.
\end{align}
Compared to (\ref{mse5_0}), the proposed approach can achieve the performance of the MMSE estimator if accurate path delay estimation and enough quantization bits when the SNR is large enough.

\section{Simulation Results}

In this section, computer simulation is conducted to verify the proposed approach. In the simulation, we consider an OFDM system with $K=256$ subcarriers and the subcarrier spacing is $15$ KHz. We adopt a typical uniform-linear-array (ULA) spaced by half wave-length. The antenna number is $M=64$ and the size of the effective channel is $D=6$. The actual CIR is composed of $L = 6$ paths. An exponential power delay profile is assumed and the path delays are uniformly distributed within $[0,\tau_{\mathrm{max}}]$  with $\tau_{\mathrm{max}}=5\mu\text{s}$. Each path has $20$ unresolvable subpaths and each subpath has a random angle of departure (AoD). In practical systems, the AoDs for different paths can be distributed within a local or a rather wide range. To take various cases into account, the AoDs are assumed independently and uniformly distributed within a range that has a random central angle and a random angular spread uniformly distributed within $[-\pi,\pi)$ and $[0,\pi/2]$, respectively.

\begin{figure}
  \centering
  \centering
  \includegraphics[width=5in]{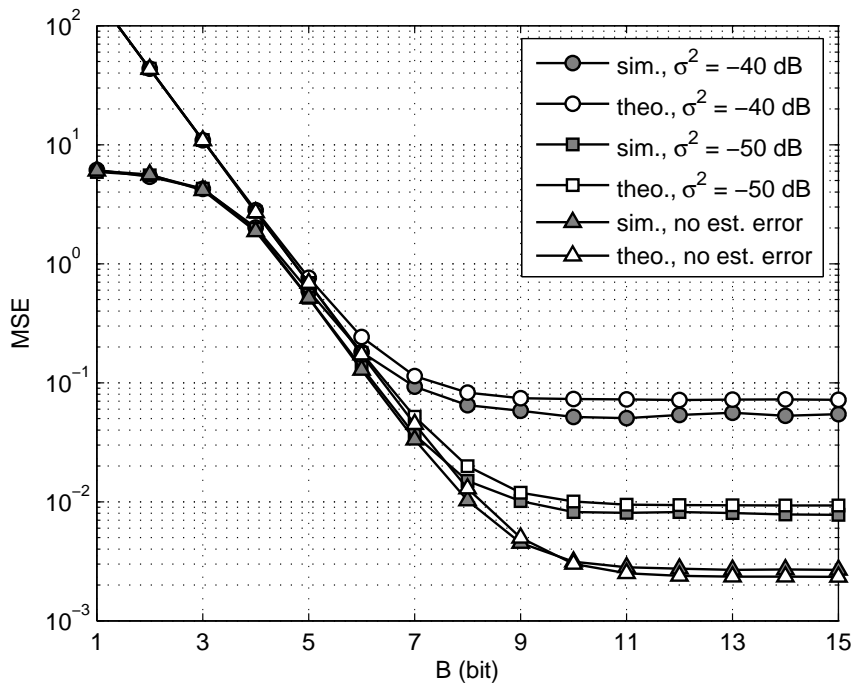}\\
  \caption{MSE versus the number of quantization bits for different variances of estimation error at SNR=$10$ dB.}\label{myfig1}
\end{figure}

Fig.~\ref{myfig1} shows the MSE versus the number of quantization bits with different variances of estimation errors normalized by $\tau_{\mathrm{max}^2}/12$. The theoretical MSE in (\ref{mse4}) is also shown for comparison. Since the theoretical MSE is derived based on the assumption that the number of quantization bits is large, we can observe that the simulated MSEs almost coincide with the theoretical ones when $B$ is large, but a gap exists when $B$ is small. For $\sigma^2=-40$ dB, the MSE cannot be improved further when $B>8$ because the estimation error is dominant in this situation. As the accuracy of the path delay estimation improves, the MSE can be further reduced by using more quantization bits.

\begin{figure}
  \centering
  \centering
  \includegraphics[width=5in]{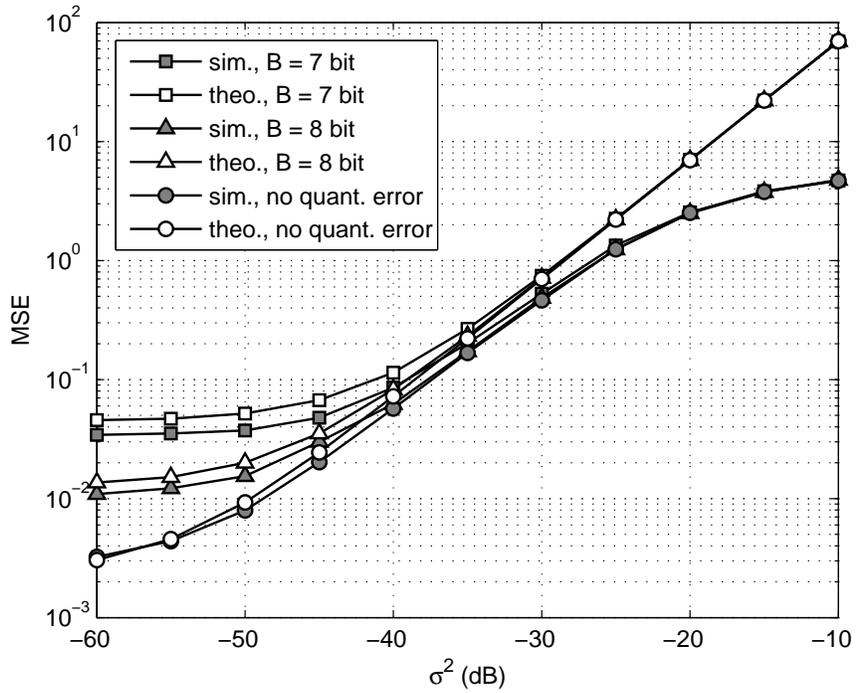}\\
  \caption{MSE versus variances of estimation error with different numbers of quantization bits at SNR=$10$ dB.}\label{myfig2}
\end{figure}

Fig.~\ref{myfig2} shows the MSE versus the variances of the estimation errors with different numbers of quantization bits. When the number of bits is small, the MSE can be hardly improved further by improving the estimation accuracy because the quantization error is dominant in this situation. The performance can be further improved by reducing the variance of estimation error when more quantization bits are used.

\begin{figure}
  \centering
  \centering
  \includegraphics[width=5in]{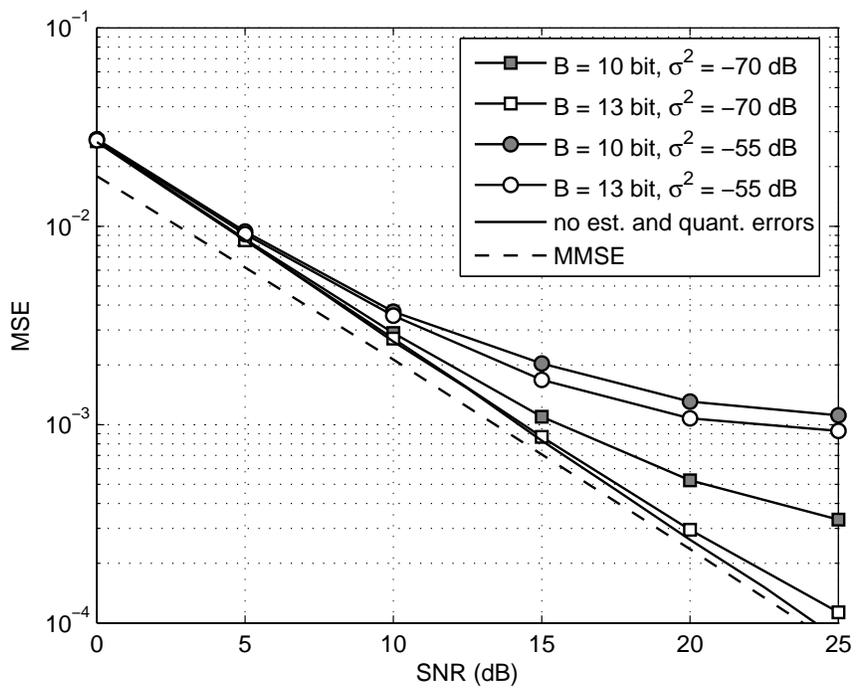}\\
  \caption{MSE versus the SNR with different quantization bits and variances of the estimation errors.}\label{myfig3}
\end{figure}

\begin{figure}
  \centering
  \centering
  \includegraphics[width=5in]{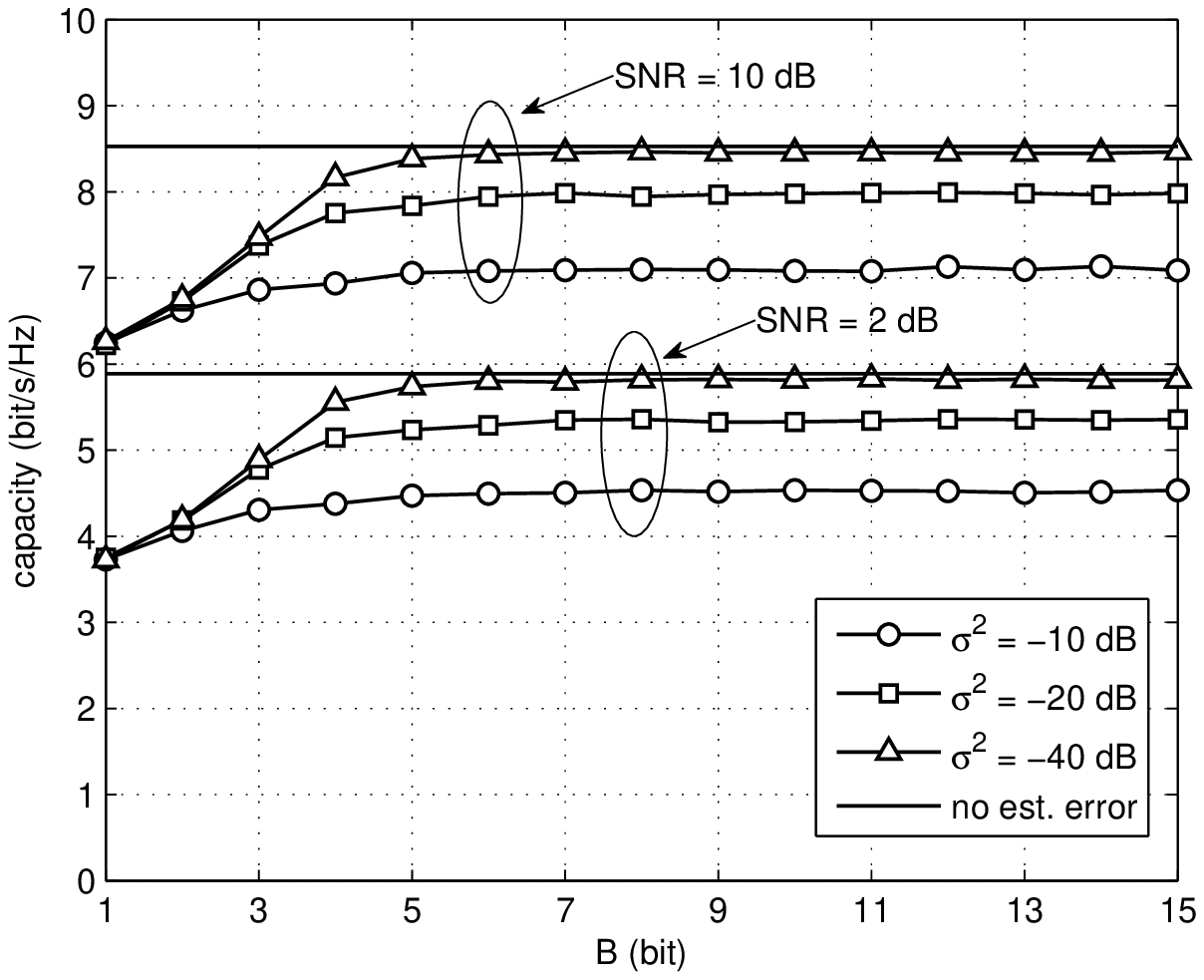}\\
  \caption{Capacity versus quantization bits with different variances of the estimation errors.}\label{myfig4}
\end{figure}

Fig.~\ref{myfig3} shows the MSE versus the SNR with different quantization bits and variances of the estimation errors. As expected, the MSE can be improved by using more quantization bits or improving the accuracy of path delay estimation. On the other hand, when the number of bits is large enough and the path delay estimation is accurate enough, the proposed approach can achieve near-MMSE performance at the high SNR domain. This coincides with our analysis in Section IV.

Fig.~\ref{myfig4} shows the capacity versus number of bits for different variances of estimation errors. In this figure, the BS uses the effective CSI fed back by the UE for downlink precoding. We assume the estimated channels at the UE can be perfectly fed back to the BS such that the performance is only affected by the channel estimation error. From the figure, only $5$ bits are enough to achieve the case with ideal effective CSI.

\begin{figure}
  \centering
  \centering
  \includegraphics[width=5in]{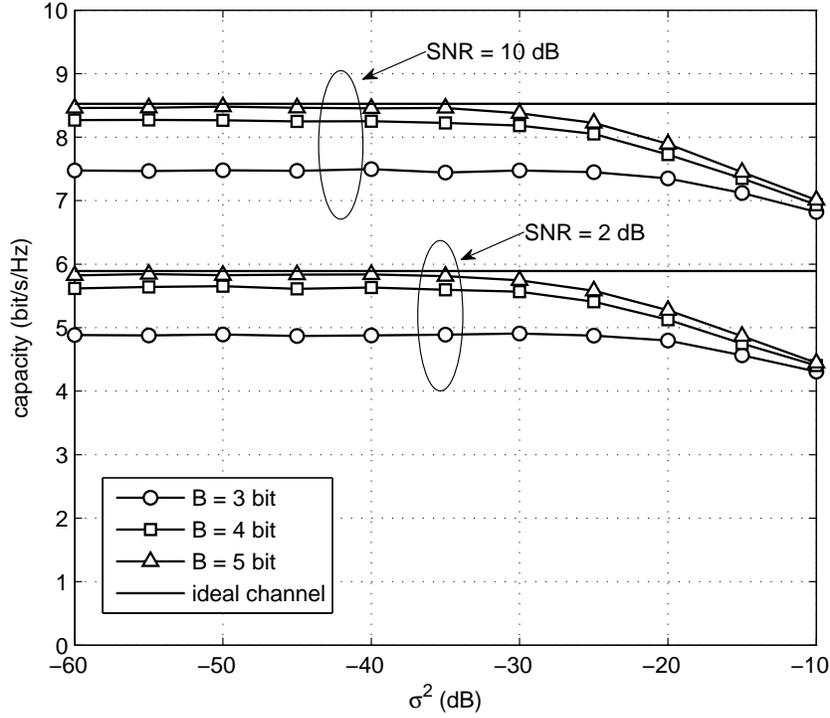}\\
  \caption{Capacity versus variances of estimation errors with different number of quantization bits.}\label{myfig5}
\end{figure}

Fig.~\ref{myfig5} shows the capacity versus variances of estimation errors with different number of quantization bits. Similar to that in Fig.~\ref{myfig4}, we still assume the estimated channels at the UE can be perfectly fed back to the BS. From the figure, the capacity can be hardly improved when $\sigma^2<-25 $ dB, which means that more accurate path delay estimation is unnecessary.
\section{Conclusions}
In this paper, we have designed and analyzed the downlink channel estimation in FDD-based massive MIMO systems with cascaded precoding. Assuming the downlink and the uplink have the same path delays, the path delays are first estimated at the BS, and then quantized and fed forward to the UE. In this case, the UE can obtain the a priori knowledge of the downlink channels. The accuracy of downlink channel estimation can be therefore significantly improved. Our simulation results have shown that the proposed approach can achieve near-MMSE performance given accurate path delay estimation and sufficient quantization bits, which also coincides with our theoretical analysis.

\appendices

\renewcommand{\theequation}{A.\arabic{equation}}
\setcounter{equation}{0}
\section{}
The MSE for channel estimation in (\ref{chan_mse}) can be rewritten in a matrix form as
\begin{align}\label{B1}
\mathrm{MSE}&=\sum_{d=0}^{D-1}\mathrm{E}(\|\widehat{\mathbf{S}}\widehat{\boldsymbol{\beta}}_d-\mathbf{S}\boldsymbol{\beta}_d\|_2^2)\nonumber\\
&=\sum_{d=0}^{D-1}\mathrm{Tr}\{\widehat{\mathbf{S}}\mathrm{E}(\widehat{\boldsymbol{\beta}}_d\widehat{\boldsymbol{\beta}}_d^{\mathrm{H}})\widehat{\mathbf{S}}^{\mathrm{H}}-\mathbf{S}\mathrm{E}(\boldsymbol{\beta}_d\widehat{\boldsymbol{\beta}}_d^{\mathrm{H}})\widehat{\mathbf{S}}^{\mathrm{H}}
-\widehat{\mathbf{S}}\mathrm{E}(\widehat{\boldsymbol{\beta}}_d\boldsymbol{\beta}_d^{\mathrm{H}})\mathbf{S}^{\mathrm{H}}+\mathbf{S}\mathrm{E}(\boldsymbol{\beta}_d\boldsymbol{\beta}_d^{\mathrm{H}})\mathbf{S}^{\mathrm{H}}\}.
\end{align}
Therefore, we need calculate $\mathrm{E}({\boldsymbol{\beta}}_d{\boldsymbol{\beta}}_d^{\mathrm{H}})$, $\mathrm{E}(\widehat{\boldsymbol{\beta}}_d\widehat{\boldsymbol{\beta}}_d^{\mathrm{H}})$, and $\mathrm{E}({\boldsymbol{\beta}}_d\widehat{\boldsymbol{\beta}}_d^{\mathrm{H}})$, respectively.\par
\subsection{Calculation of $\mathrm{E}({\boldsymbol{\beta}}_d{\boldsymbol{\beta}}_d^{\mathrm{H}})$}
From (\ref{beta_exp}), the correlation function of $\beta_d[l]$ can be given by
\begin{align}\label{corr_beta}
\mathrm{E}(\beta_{d}[l]\beta_{d}^*[l_1])=\mathbf{u}_s^{\mathrm{H}}[d]\mathbf{R}_{s,l}\mathbf{u}_s[d]\delta[l-l_1].
\end{align}
Therefore, $\mathbf{R}_{\beta,d}\triangleq\mathrm{E}({\boldsymbol{\beta}}_d{\boldsymbol{\beta}}_d^{\mathrm{H}})$ is a diagonal matrix with the $(l,l)$-th entry
\begin{align}\label{B1_0}
[\mathbf{R}_{\beta,d}]_{(l,l)}=\mathbf{u}_s^{\mathrm{H}}[d]\mathbf{R}_{s,l}\mathbf{u}_s[d].
\end{align}\par

\subsection{Calculation of $\mathrm{E}(\widehat{\boldsymbol{\beta}}_d\widehat{\boldsymbol{\beta}}_d^{\mathrm{H}})$ and $\mathrm{E}({\boldsymbol{\beta}}_d\widehat{\boldsymbol{\beta}}_d^{\mathrm{H}})$}
For the calculation of $\mathrm{E}(\widehat{\boldsymbol{\beta}}_d\widehat{\boldsymbol{\beta}}_d^{\mathrm{H}})$ and $\mathrm{E}({\boldsymbol{\beta}}_d\widehat{\boldsymbol{\beta}}_d^{\mathrm{H}})$, from (\ref{LS}), we have
\begin{align}\label{B2}
\widehat{\boldsymbol{\beta}}=\left(\frac{1}{K}\widehat{\mathbf{X}}^{\mathrm{H}}\widehat{\mathbf{X}}\right)^{-1}\frac{1}{K}\widehat{\mathbf{X}}^{\mathrm{H}}\mathbf{X}\boldsymbol{\beta}+\left(\frac{1}{K}\widehat{\mathbf{X}}^{\mathrm{H}}\widehat{\mathbf{X}}\right)^{-1}\frac{1}{K}\widehat{\mathbf{X}}^{\mathrm{H}}\mathbf{z}.
\end{align}\par
\subsubsection{Calculation of $\frac{1}{K}\widehat{\mathbf{X}}^{\mathrm{H}}\widehat{\mathbf{X}}$}
Note that
\begin{align}\label{B3}
\frac{1}{K}\widehat{\mathbf{X}}^{\mathrm{H}}\widehat{\mathbf{X}}=\left(
\begin{array}{ccc}
  \frac{1}{K}\widehat{\mathbf{S}}^{\mathrm{H}}\widehat{\mathbf{S}} & \cdots &  \frac{1}{K}\widehat{\mathbf{S}}^{\mathrm{H}}\mathbf{A}_0^{\mathrm{H}}\mathbf{A}_{D-1}\widehat{\mathbf{S}}\\
  \vdots & \ddots & \vdots \\
  \frac{1}{K}\widehat{\mathbf{S}}^{\mathrm{H}}\mathbf{A}_{D-1}^{\mathrm{H}}\mathbf{A}_{0}\widehat{\mathbf{S}} & \cdots & \frac{1}{K}\widehat{\mathbf{S}}^{\mathrm{H}}\widehat{\mathbf{S}}
\end{array}
\right)
\end{align}\par
For the diagonal submatrices in (\ref{B3}), we have
\begin{align}\label{B3_0}
\frac{1}{K}\widehat{\mathbf{S}}^{\mathrm{H}}\widehat{\mathbf{S}}=\mathbf{I},
\end{align}
where we have used the identity in (\ref{orth}).\par

For the $(d_1,d_2)$-th off-diagonal submatrix ($d_1\neq d_2$) in (\ref{B3}),
\begin{align}\label{B4}
\frac{1}{K}\widehat{\mathbf{S}}^{\mathrm{H}}\mathbf{A}_{d_1}^{\mathrm{H}}\mathbf{A}_{d_2}\widehat{\mathbf{S}}=\frac{1}{K}\left[
\begin{array}{c}
  \mathbf{s}^{\mathrm{H}}(\widehat{\tau}_0)\mathbf{A}_{d_1}^{\mathrm{H}} \\
  \vdots \\
  \mathbf{s}^{\mathrm{H}}(\widehat{\tau}_{L-1})\mathbf{A}_{d_1}^{\mathrm{H}}
\end{array}
\right][\mathbf{A}_{d_2}\mathbf{s}(\widehat{\tau}_0),\cdots,\mathbf{A}_{d_2}\mathbf{s}(\widehat{\tau}_{L-1})],
\end{align}
the $(l_1,l_2)$-th entry is given by
\begin{align}\label{B5}
\frac{1}{K}\mathbf{s}^{\mathrm{H}}(\widehat{\tau}_{l_1})\mathbf{A}_{d_1}^{\mathrm{H}}\mathbf{A}_{d_2}\mathbf{s}(\widehat{\tau}_{l_2})=\frac{1}{K}\sum_{k=0}^{K-1}e^{j\varphi[k]},
\end{align}
where $\varphi[k]=\phi[k]+\frac{2\pi k}{T}(\widehat{\tau}_{l_1}-\widehat{\tau}_{l_2})$ with $\phi[k]=\phi_{d_2}[k]-\phi_{d_1}[k]$. Since $\phi_{d}[k]$'s are uniformly distributed within $[-\pi,\pi)$ and mutually independent for different subcarriers and different eigen-beams, $\phi[k]$'s with $k=0,1,\cdots,K-1$ can be viewed as independently identically distributed random variables with zero means and a common probability density function (pdf)
\begin{align}\label{B7}
p_X(x)=
\begin{cases}
\displaystyle{\frac{1}{2\pi}-\frac{|x|}{4\pi^2}} & |x|<2\pi\\
0 & \mathrm{otherwise}
\end{cases}.
\end{align}
Accordingly, $\varphi[k]$'s with $k=0,1,\cdots,K-1$ are also independently distributed random variables but with mean $\mathrm{E}(\varphi[k])=\frac{2\pi k}{T}(\widehat{\tau}_{l_1}-\widehat{\tau}_{l_2})$. Since the functions of independent random variables are still independent, $e^{j\varphi[k]}$'s are therefore independent for different subcarriers. Then, following the law of large numbers, we have
\begin{align}\label{B6}
\frac{1}{K}\sum_{k=0}^{K-1}e^{j\varphi[k]}&=\frac{1}{K}\sum_{k=0}^{K-1}\mathrm{E}(e^{j\varphi[k]})\nonumber\\
&=\frac{1}{K}\sum_{k=0}^{K-1}e^{j\frac{2\pi k}{T}(\widehat{\tau}_{l_1}-\widehat{\tau}_{l_2})}\mathrm{E}(e^{j\phi[k]}).
\end{align}
Using the pdf in (\ref{B7}), we have
\begin{align}\label{B8}
\mathrm{E}(e^{j\phi[k]})=\int_{-2\pi}^{2\pi}\left(\frac{1}{2\pi}-\frac{|\phi[k]|}{4\pi^2}\right)e^{j\phi[k]}d\phi[k]=0.
\end{align}
Substituting (\ref{B8}) into (\ref{B6}), we obtain
\begin{align}
\frac{1}{K}\mathbf{s}^{\mathrm{H}}(\widehat{\tau}_{l_1})\mathbf{A}_{d_1}^{\mathrm{H}}\mathbf{A}_{d_2}\mathbf{s}(\widehat{\tau}_{l_2})=\frac{1}{K}\sum_{k=0}^{K-1}e^{j\varphi[k]}=0,
\end{align}
and therefore,
\begin{align}
\frac{1}{K}\widehat{\mathbf{S}}^{\mathrm{H}}\mathbf{A}_{d_1}^{\mathrm{H}}\mathbf{A}_{d_2}\widehat{\mathbf{S}}=\mathbf{0},
\end{align}
for $d_1\neq d_2$. In other words, the off-diagonal submatrices in (\ref{B3}) are all zeros when the number of subcarriers is large enough.\par
As a result,
\begin{align}\label{B9}
\frac{1}{K}\widehat{\mathbf{X}}^{\mathrm{H}}\widehat{\mathbf{X}}=\mathbf{I}.
\end{align}
\subsubsection{Calculation of $\frac{1}{K}\widehat{\mathbf{X}}^{\mathrm{H}}\mathbf{X}$}
Similar to the derivation above, we can obtain that
\begin{align}
\frac{1}{K}\widehat{\mathbf{X}}^{\mathrm{H}}{\mathbf{X}}=\left(
\begin{array}{ccc}
  \frac{1}{K}\widehat{\mathbf{S}}^{\mathrm{H}}{\mathbf{S}} &  &  \\
   & \ddots &  \\
   &  & \frac{1}{K}\widehat{\mathbf{S}}^{\mathrm{H}}{\mathbf{S}}
\end{array}
\right),
\end{align}
since the off-diagonal submatrices are zeros. If we assume the quantized delays, $\widehat{\tau}_l$'s, are close to the real delays, $\tau_l$'s, then $\frac{1}{K}\widehat{\mathbf{S}}^{\mathrm{H}}{\mathbf{S}}$ can be approximated by
\begin{align}
\frac{1}{K}\widehat{\mathbf{S}}^{\mathrm{H}}{\mathbf{S}}=\mathbf{\Lambda},
\end{align}
where $\mathbf{\Lambda}$ is an $L\times L$ diagonal matrix with the $(l,l)$-th entry
\begin{align}
[\mathbf{\Lambda}]_{(l,l)}=\mathrm{sinc}\left[\frac{\pi(\tau_l-\widehat{\tau}_l)K}{T}\right]e^{j\frac{\pi (K-1)(\tau_l-\widehat{\tau}_l)}{T}},
\end{align}
and therefore
\begin{align}\label{B13}
\frac{1}{K}\widehat{\mathbf{X}}^{\mathrm{H}}{\mathbf{X}}=\left(
\begin{array}{ccc}
  \mathbf{\Lambda} &  &  \\
   & \ddots &  \\
   &  & \mathbf{\Lambda}
\end{array}
\right).
\end{align}
\par
Using (\ref{B9}) and (\ref{B13}), (\ref{B2}) can be simplified as
\begin{align}\label{B15}
\widehat{\boldsymbol{\beta}}_d=\mathbf{\Lambda}\boldsymbol{\beta}_d+\frac{1}{K}\mathbf{S}^{\mathrm{H}}\mathbf{A}_d\mathbf{z}.
\end{align}
It is therefore easy to obtain that
\begin{align}
&\mathrm{E}(\widehat{\boldsymbol{\beta}}_d\widehat{\boldsymbol{\beta}}_d^{\mathrm{H}})=\mathbf{\Lambda}\mathbf{R}_{\beta,d}\mathbf{\Lambda}^{\mathrm{H}}+\frac{N_0}{K}\mathbf{I},\label{B17}\\
&\mathrm{E}({\boldsymbol{\beta}}_d\widehat{\boldsymbol{\beta}}_d^{\mathrm{H}})=\mathbf{R}_{\beta,d}\mathbf{\Lambda}^{\mathrm{H}}.\label{B18}
\end{align}\par
Using (\ref{B1_0}), (\ref{B17}) and (\ref{B18}), (\ref{B1}) can be rewritten as
\begin{align}
\mathrm{MSE}&=\sum_{d=0}^{D-1}\mathrm{Tr}\left\{\widehat{\mathbf{S}}\left(\mathbf{\Lambda}\mathbf{R}_{\beta,d}\mathbf{\Lambda}^{\mathrm{H}}+\frac{N_0}{K}\mathbf{I}\right)\widehat{\mathbf{S}}^{\mathrm{H}}-\mathbf{S}\mathbf{R}_{\beta,d}\mathbf{\Lambda}^{\mathrm{H}}\widehat{\mathbf{S}}^{\mathrm{H}}
-\widehat{\mathbf{S}}\mathbf{\Lambda}\mathbf{R}_{\beta,d}\mathbf{S}^{\mathrm{H}}+\mathbf{S}\mathbf{R}_{\beta,d}\mathbf{S}^{\mathrm{H}}\right\}\nonumber\\
&=\sum_{d=0}^{D-1}\left\{K\sum_{l=0}^{L-1}[\mathbf{R}_{\beta,d}]_{(l,l)}(1-|[\mathbf{\Lambda}]_{(l,l)}|^2)+N_0 L\right\}\nonumber\\
&=K\sum_{l=0}^{L-1}\sum_{d=0}^{D-1}[\mathbf{R}_{\beta,d}]_{(l,l)}\left\{1-\mathrm{sinc}^2\left[\frac{\pi (\widehat{\tau}_{l}-\tau_l)K}{T}\right]\right\}+N_0 LD.
\end{align}
From (\ref{B1_0}), we have
\begin{align}
\sum_{d=0}^{D-1}[\mathbf{R}_{\beta,d}]_{(l,l)}=\mathrm{Tr}\{\mathbf{U}_s^{\mathrm{H}}\mathbf{R}_{s,l}\mathbf{U}_s\}.
\end{align}
As a result, the MSE of channel estimation can be finally obtained as
\begin{align}
\mathrm{MSE}=K\sum_{l=0}^{L-1}\mathrm{Tr}\{\mathbf{U}_s^{\mathrm{H}}\mathbf{R}_{s,l}\mathbf{U}_s\}\left\{1-\mathrm{sinc}^2\left[\frac{\pi (\widehat{\tau}_{l}-\tau_l)K}{T}\right]\right\}+N_0 LD.
\end{align}

\renewcommand{\theequation}{B.\arabic{equation}}
\setcounter{equation}{0}
\section{}
Replace $\mathbf{A}^{\mathrm{H}}\mathbf{A}$ with $\mathrm{E}(\mathbf{A}^{\mathrm{H}}\mathbf{A})=\mathbf{I}$, then, similar to the derivation in \cite{OEdfors}, the MSE in (\ref{mse_mmse}) can be expressed by
\begin{align}\label{C1}
\mathrm{MSE}_{\mathrm{MMSE}}=N_0\sum_{i=0}^{DK-1}\frac{\lambda_b[i]}{\lambda_b[i]+N_0}.
\end{align}
From (\ref{CFR_CIR}), we have $\mathbf{b}_d=\mathbf{S}\boldsymbol{\beta}_d$ and thus $\mathbf{R}_b$ can be rewritten as
\begin{align}\label{C2}
\mathbf{R}_b=(\mathbf{I}\otimes\mathbf{S})\mathbf{R}_{\beta}(\mathbf{I}\otimes\mathbf{S}^{\mathrm{H}}),
\end{align}
where $\otimes$ denotes the Kronecker product and
\begin{align}\label{C3}
\mathbf{R}_{\beta}=\left(
\begin{array}{ccc}
  \mathrm{E}(\boldsymbol{\beta}_{0}\boldsymbol{\beta}_{0}^{\mathrm{H}}) & \cdots & \mathrm{E}(\boldsymbol{\beta}_0\boldsymbol{\beta}_{D-1}^{\mathrm{H}}) \\
  \vdots & \ddots & \vdots \\
  \mathrm{E}(\boldsymbol{\beta}_{D-1}\boldsymbol{\beta}_{0}^{\mathrm{H}}) & \cdots &  \mathrm{E}(\boldsymbol{\beta}_{D-1}\boldsymbol{\beta}_{D-1}^{\mathrm{H}})
\end{array}
\right).
\end{align}
From Appendix A, $\mathrm{E}(\boldsymbol{\beta}_{d_1}\boldsymbol{\beta}_{d_2}^{\mathrm{H}})$ is a diagonal matrix and thus $\mathrm{rank}\{\mathrm{E}(\boldsymbol{\beta}_{d_1}\boldsymbol{\beta}_{d_2}^{\mathrm{H}})\}=L$. Accordingly,
\begin{align}\label{C4}
\mathrm{rank}\{\mathbf{R}_{\beta}\}=DL
\end{align}
Although it is in general difficult to prove (\ref{C4}) in a strict sense, our numerical results show that (\ref{C4}) always holds for different situations. Actually, rank deficiency of $\mathbf{R}_{\beta}$ is a very strong condition that can be hardly achieved in practical engineering problems, and therefore $\mathbf{R}_{\beta}$ is always with full rank.\par

When the number of subcarriers is very large, the relation in (\ref{orth}) means that the columns of $\mathbf{S}$ are mutually orthogonal and therefore $\mathbf{S}$ is full column rank matrix. As a result, we can obtain from (\ref{C2}) that
\begin{align}
\mathrm{rank}\{\mathbf{R}_b\}=\mathrm{rank}\{\mathbf{R}_{\beta}\}=DL.
\end{align}
In other words, there are only $DL$ significant eigenvalues for $\mathbf{R}_b$ while the others are very small and thus can be omitted. As a result, the MSE in (\ref{C1}) is reduced to
\begin{align}
\mathrm{MSE}_{\mathrm{MMSE}}=N_0\sum_{i=0}^{DL-1}\frac{\lambda_b[i]}{\lambda_b[i]+N_0},
\end{align}
which is exactly (\ref{mse_mmse2}).

\bibliographystyle{IEEEtran}
\bibliography{IEEEabrv,lsmimobib}

\end{document}